# Self Managed Security Cell, a security model for the Internet of Things and Services

Pierre de Leusse*/**, Panos Periorellis[1], Theo Dimitrakos** and Srijith K. Nair**
*Newcastle Univeristy
**British Telecom
Pierre.de-leusse@ncl.ac.uk

*Abstract*—The Internet of Things and Services is a rapidly growing concept that illustrates that the ever increasing amount of physical items of our daily life which become addressable through a network could be made more easily manageable and usable through the use of Services. This surge of exposed resources along with the level of privacy and value of the information they hold, together with the increase of their usage make for an augmentation in the number of the security threats and violation attempts that existing security systems do not appear robust enough to address. In this paper, the authors underline this increase in risk and identify the requirements for resources to be more resilient in this type of environment while keeping an important level of flexibility. In addition, the authors propose an architectural model of Self Managed Security Cell, which leverages on current knowledge in large scale security systems, information management and autonomous systems.

*Index Terms*— Internet of Things, Security, Service Oriented Architecture

## I. INTRODUCTION

The concept of the Internet of Things and Services (IoT&S) is based on the possibility of seamless integration of physical objects such as sensors or home appliances (i.e. things) and services, which can be loosely defined as a network interface that exposes a piece of functionality.

The IoT&S is gaining momentum in the academic and industrial areas alike [1][2]. Many key enabling technologies such as middleware and sensor networking are now available, have gained maturity and their usage is expected to become more and more common. The IoT&S involves and leverages on current knowledge in automatic identification and communication of things and services.

However, the envisioned world of pervasive computing would be closer to realisation if the embedded devices were able to communicate and interoperate with each other so as to cooperate.

The development of this paradigm has indeed created new types of requirements that can be expressed in terms of the quantity of resources connected, density of the connections and complexity to manage both elements together. The IoT&S encompasses that more and more resources (e.g. applications, devices) are exposed over the network (e.g. Internet, domestic, corporate, etc). These resources are linked in groups (e.g. federations) where one resource can be part of several clusters. For instance, one mobile phone device is in the user's domestic systems (with personal laptop, mobile, car, etc) and in his company's (with other corporate "things" such as computers, servers, printers, etc.) if this one has been provided by it. This means that more than one entity can administrate the resources, at times only in specific context, and sets policies for them. In term of management, some of these policies might be relevant only for a specific resource, in a particular context, some others relevant for different resources in the same federation. For instance a spam filter of an email box and contacts details in messaging tools could be reused by the same user to filter calls as well as texts on his mobile.

In this paper, the authors focus on the need to define and comprehend the requirements that have risen with the IoT&S in order to model a generic architectural framework to secure resources. The goal of this work is not to define yet another security protocol or grammar, but to attempt to identify an architecture that leverages on the existing knowledge in large scale security systems, information management and autonomous systems. This model underlines the need for interoperability, decentralisation, automation and contextualisation in modern security systems.

First, let us define a resource as an asset that is exposed on a network. Typical assets are applications or devices. The first type covers a wide range from data service or email software, to business application. Devices are becoming an increasingly important part of network as technologies develop. Indeed, current mobile devices are now equipped with a network interface, the use of sensor is also spreading and in a near future it seems unavoidable that cars, fridges and other everyday tools will follow.

In the first section, the objectives of this paper are the established, then the specific requirements for security in the IoT&S are listed and related works are introduced. Following this, a model that takes into account the previous discussion is presented and a potential implementation of it is shortly introduced and discussed.

---

[1] Panos recently moved to Microsoft Corp. And can be contacted at Panos.Periorellis@microsoft.com

47

## II. OBJECTIVES

We believe that such issues can be resolved by developing an infrastructure that manages all communications between resources. We want to investigate and develop security mechanisms as part of a third party infrastructure or dedicated layer that mediates communications between resources while at the same time it does not challenge the organisational policies of those services.

Our view is that the infrastructure can exist as a set of managed components, ideally when possible services, that users can make use of in order to dependably expose a resource on a network. We would like to avoid imposing any additional requirements at the resource side by offering flexibility as part of the infrastructure that would not compromise aspects of the resource. Numerous solutions seem to be tailored to particular domains that require some degree of trade off between those important non-functional properties. We want to offer such flexibility at the level of the infrastructure in terms of message transfer protocols that do not compromise the policies of those resources. The aim is to enable resources to be network ready without any prerequisites that challenge their autonomy. The added value of the infrastructure is that it enables the use of resources without imposing any architectural constraints and at the same time maintaining the responsibility for the delivery of a message and its support of certain attributes. Given the earlier discussion these guarantees may relate to fairness, audit, anonymity and privacy, but we are also looking to extend or complement these with mechanisms pricing and others as presented by the authors in [3].

## III. REQUIREMENTS FOR A SECURITY CELL

The principal ideas presented in this paper are that in order to deal with and improve the security of resources in large scale distributed systems there is a strong need to address the following requirements:

### A. Interoperability

The interoperability issue in the security of IoT&S can be separated in three main domains. The first addresses the semantic of communication, the second the grammar of communication and finally the third regards the operational connection. This last point will be developed in the next section on automation.

In order for resources to understand each other, the first factor is to be able to know what the data exchanged means. For instance, two antivirus software from different brands are not capable of using each other's security patches and updates. This issue, when applied to an even wider audience (e.g. anti spam together with firewall antivirus), increased the complexity even more. Two types of mechanisms have been created to address this, translation protocols and common standardised semantics.

 -- Security ontologies and translation mechanisms have been developed in order for different firewall and anti virus software to exchange data [4][5][6].

If resources can understand the semantic of the data they exchange, they do not necessarily express the same fact using a shared grammar. In SOA, a field very active in the domain of interoperability, several grammars have been developed and are used for security related issues:

 -- eXtensible Access Control Markup Language (XACML) [7][8] is an XML based OASIS Web service oriented standard for communicating access control policies between services.
 -- Security Assertion Markup Language (SAML) [9][10][11] standard is an OASIS standardised specification for expressing, requesting and delivering assertions regarding the credentials of various entities (users, computers, printers etc). Shibboleth is well known implementation of SAML, implemented on the OpenSAML [12] APIs.
 -- SecPAL is a policy and token authoring language developed by Microsoft Corporation [13]. It combines access control policies and security tokens under the same grammar which in turn conforms to a formal model. An XML schema for serializing SecPAL policies and tokens into XML.

The aforementioned grammars could benefit the IoT&S by providing a standard access control policy language replacing dozens of application-specific languages such as [14][15][16][17][18]. The interest with XACML and SecPAL is that they allow answering both issues of securing the content exchanged and the communication interfaces. Additionally the potential of providing semantic translators between grammars should also be examined.

### B. Automation

The automation of the security tasks is of paramount importance in the IoT&S. Indeed, as the amount of resources and their level of associations increase, matched with the potential augmentation in usage and the fact that these resources are exposed on a network, it becomes less and less viable for a human based management, at least at the single device level. The lack of automation can create two types of issue regarding detection of threat and action based upon the detection. The first one being that all resources cannot be manually made aware of all new threats as these evolve and new are created. The second one regards the decisions and actions that have to be taken upon detection of a threat. If we assume that interoperability has been achieved (c.f. previous chapter on interoperability), then resources are capable of understanding each other's security related information. What automation promotes is a more advanced interoperability regarding the exposure of management type interfaces which allow exchanging data and permit resources to know how to manipulate each other – when allowed – in order to take actions. Two main standards have been developed in order to allow this:

 -- Web Services Distributed Management (WS-DM) [19], an OASIS standard, tackles simplifying the management of heterogeneous IT resources. WSDM defines a standard way of how to represent and access the management interfaces.

-- Still in the SOA domain, WS-Management (WS-M) [20] is a specification of a SOAP-based protocol for the management of servers, devices and applications. It provides a common way for systems to access and exchange management information across the IT infrastructure.

*C. Decentralisation*

Many resources maintain their own policy stores, typical examples are email clients which allow creating and managing their own rules. For practical reasons it is therefore unavoidable to manage decentralised information store and decision making. Additionally, for security reasons it is best not to duplicate and share security related data outside of its context.

*D. Contextualisation*

Resources can be required to offer different functions and different types of data according to the situation. The faculty to adequately segregate context without allowing them to overlap is a key requirement in the IoT&S. Additionally, it might be unavoidable for some resources to be dependant of different types and sources of management in different contexts (c.f. mobile device scenario in the introduction).

Practically, these requirements mean that instead of an administrator reading logs sent by different security related systems (e.g. anti spam, firewall, anti virus) and making decisions upon the understanding and use of this information, we would need the relationship between receiving the information and acting automated. In addition, due to the dynamic nature and large scale of the Internet, resources should be able to automatically discover what other resources it might be relevant to pull data from and push to. This would help increasing the dynamicity of the updates and potentially improving the security by improving the general security related awareness. Finally, a resource should be made aware of the potentiality of interacting in different environments under different constraints.

## IV. RELATED WORK

The notion IoT&S encompasses a wider range of devices that use TCP/IP and protocols on top of that to communicate with PC's or other devices. Mobile phones, car navigators, handheld cameras are but a few embedded devices that that are part of this notion. A lot of these devices use web service technology as a means to communicate or synchronise themselves with other 'things' on the internet. Apart from typical consumer products the notion of IoT&S has extended to manufacturing and industrial automation. Factory shop floors are typically filled with sensors, robots, and other smart embedded devices. Typically a production line works in isolation according to some ERP (Enterprise Resource Planning) specification. Getting production line embedded devices to talk in real time to ERP systems can be beneficial. Technologies such as the ones that consumer devices use are the obvious one to be tried first. The need for developing web services type interfaces for shop floor embedded devices is being currently addressed in [21]. Web services are used to wrap part of the devices' functionality in order to allow bidirectional communication between back end ERP services. The reason can vary between simple auditing and monitoring to dynamic reconfiguration and troubleshooting. SOA devices are also becoming popular [22], being able to expose the interface of part of the functionality of a device can be beneficial.

Device Profile for Web services is an XML specification for defining standard APIs for such devices. Although in its infancy it is currently assessed in car production lines and other domains [23]. In an area where SAP mainly dominated with XMii and their set of proprietary protocols Web services and the related specs provided a sense of freedom between rigid devices (shop floor) and the internet.

Self-Managed Cell (SMC) [24], a policy-based architecture that integrates services, managed resources and a policy interpreter by means of an event bus is used in [25] to manage body sensor networks. Body sensor networks consist of on-body wireless sensors attached to patients to monitor the health and well being. Such systems need to adapt autonomously to changes in context, user activity, device failure, and the availability or loss of services.

The notion of self managed cells is also interested and it would provide an extension to the current state of research in the area of self managed embedded devices. Karnouskos [26] expressed the need to self managed devices that are aware of their state (hence being able to transmit it) as well as their interface (so that other devices can link to them). The complete self management life cycle of a device would be complemented using a set of protocols and configuration interfaces that would enable devices to self 'heal' or configure dynamically according to the collective state reached by the shop floor. Current research and market analysis shows that smart devices are important in the not too far future. Many large IT companies are moving in areas such as industrial automation, house automation, as well as embedded device integration and research areas where issues such as self management and autonomy are addressed are important.

## V. ANATOMY OF A SECURITY CELL

In [27], a pattern of Self-Managed Cells (SMC) that can be federated is defined. In the following chapter we leverage on this pattern to define a model of Self Managed Security Cells (SMSC).

The SMC model was designed to be able to configure itself with little or no user input and to adapt autonomously to changes. SMCs are typically composed of Policy, Discovery and Role services which through an event bus allow managing a resource through measurement and control adapters. SMC is therefore a strong candidate for the IoT&S as it takes into account the strong need for automation and autonomic. The model was however not designed to deal with security and moreover to leverage on the scale of the network to attend to improve it. The policy service is also inadequate as it only targets adaptation strategies. The SMSC model adds security and management oriented components to the SMC in order to render it capable of securing communications and behavior of the protected resource(s) with potentially the capacity to leverage on the amount of resources federated to improve the security of the network.

This refined model takes into account the requirements previously introduced in this paper in order to propose a scalable security enhancement system for distributed resources. It is however noticeable that in the case of the SMSC the system is no longer aimed at being fully autonomous as user input is expected (e.g. user add excluded domain by flagging email as spam). Another key point is that these services, for performance or business reasons, could be centralised for a set of resources just as they could be placed on the resource.

Figure 1 "Self Managed Security Cell" presents a logical view of the model.

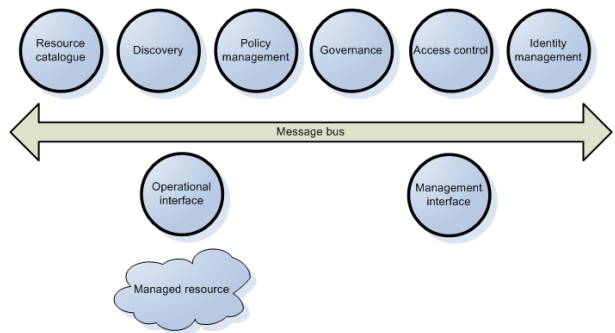

Figure 1. Self Managed Security Cell

The responsibilities of the core services shown in Figure 1 are described below.

### A. Message bus

The role of the message bus is to provide connectivity to the different parts of the SMSC. This can require providing different options such as synchronous and asynchronous communication. A complete model of such middleware is described in [28].

### B. Discovery

In order to improve the knowledge of one cell, it is essential to discover other components that could improve it. In a non centralised system this is best left to each component to potentially advertise itself, let resources register one to another when relevant or possibly discovering partners trough managed registries (e.g. UDDI). The discovery service's aims are to discover new partners and to allow the managed resource to get to know them (e.g. what they do, who/what manage them, etc).

### C. Resource catalogue

The resource catalogue allows holding the information about other cells, the different policies (e.g. access control, adaptation, etc) and configuration files necessary for the managed resource.

### D. Policy management

Given the plethora of policies in conjunction with the fact that there is not a single authority that governs these policies (the majority will stem from user requirements on how they want to protect their resources) validation is needed to make sure that there are no logical inconsistencies.

### E. Access control and Identity management

In earlier publications [29] we discussed the advantages and disadvantages of access control models, ranging from active control lists to role and task based systems. We concluded that grid computing requires infrastructure that deals with dynamic rights activation and deactivation and more importantly delegation. Cloud computing also brings together a number of resources accessible via standard web protocols. This reinforces well documented requirements such as the need for granularity, single sign on and federation but it also exposes some new ones.

Delegation is important and in particular in grid computing. Several authors have elaborated on this issue [30][31][32]. The reason being that given the sensitivity of the information that may be shared in some grid environments (which raises concerns regarding competitive advantage) parties are not expected to be assigned a single set of rights (held by roles) that would last throughout the life time of a domain. It is more likely that limited or gradual access to their resources would be granted. In order to support such dynamic behaviour a language the handles delegation models with ease is desirable.

Usability is also a major issue. Unfortunately a number of tools that offer similar services or target the same domain have failed to connect with developers and policy writers. We believe unnecessary complexity and poor usability play a major role in the success and consequent adaptation of such as language. Another criticism is that of the lack of clear semantics in Grid security tools. Our investigation in systems such as Permis and XACML [29] has shown that these systems lack any formal semantics of their constructs with the XML schema being the corner stone of their development. We need to address this and come up with a solution that combines all security related information under the same semantics. SecPAL has taken some steps towards this direction. Uniting both policies and security tokens has also been desirable in Grid environments. There have been attempts for example to combine XACML policies and SAML tokens at the XML level but have failed to address this issue at the level of semantics.

### F. Governance

This service, together with the policy management insures that the incoming security "updates" and policies have a safe impact of the current policies. In addition, it tries to find out how current changes in the sate of the resource it manages could have an effect on known resources registered in the Resource Registry. Finally, it also allows exposing the resource and its data in different contexts, according to access policies and profiles as introduced in. More details on this precise component will be given in a future article in a special issue on SOI of the British Telecom Technology Journal (BTTJ).

*G. Management interface*

The management interface allows remote configuration and management of the resource. As previously introduce, this type of interface is exposed according to certain standard to allow for interoperability and automation.

*H. Operational interface*

The operational interface is an access layer to the function of the resource itself.

The SMC model [27] is appealing in this context as it takes into account the requirements of automation as well as decentralisation, allows managing many different types of resources and is extendable enough to tolerate the integration of components supporting the other requirements.

## VI. SERVICE ORIENTED INFRASTRUCTURE – SERVICE SECURITY GATEWAY (SOI-SSG)

In the following part, the SOI-SSG [34], its main components and their roles are briefly introduced. In addition, a mapping with the Security Cell architecture previously presented is made in order to demonstrate the feasibility and practicality of the SMSC.

Figure 2 provides an overview of the security capabilities necessary for a typical Service Oriented Enterprise (SOE). In this section we focus on a core subset of the common capabilities that are necessary for a secure SOA realisation.

The first capability focuses on the problems of protecting the exposure and availability of services to the network, and ensuring confidentiality, integrity, and accountability in their end-to-end interactions.

An identity management component, the Security Token Service (STS) addresses the problems of identity brokerage federation and management of the life-cycle of circles-of-trust between identity brokers as well as the life-cycle of virtual identities and other security assertions that may be used in B2B collaborations.

An access control capability, the Policy Decision Point (PDP) which focal points are service-level usage and access control in complex, multi-administrative environments is also used.

A governance and policy management capability that allows for a safe contextualisation and usage of distributed policies is proposed.

The requirements forming the basis of these elements have been elicited by studying the business and technological requirements of a large number of business cases and pilots in research projects such as TrustCoM [35][36] and BEinGRID [37][38].

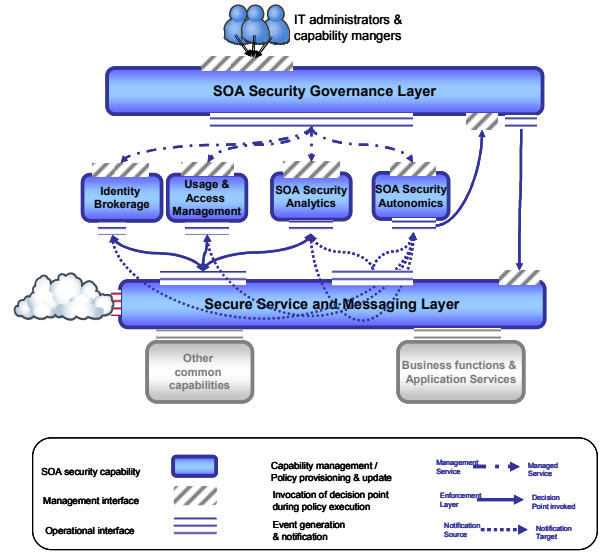

Figure 2. Overview of the SOI-SSG capabilities.

## VII. SOI-SSG MODEL ON SECURE CELL

This section brings the Secure Cell architecture in the heart of the SOI-SSG. The SSG is not targeted at managing particular "things" and therefore does not comport any operational or management interface serving this purpose. In addition, potential partners are required to register themselves through a federation [34], as the SOI-SSG does not aim at handling this level of management. The rest of the components however are direct matches and do closely conform to the Secure Cell architecture and guidelines presented in the previous sections.

The resource catalog in the SOI-SSG can be found in two locations depending on its purpose. The collaboration management layer can maintain white pages / UDDI where services are made publicly available. In addition, each partner's SOI-SSG governance layer also maintains an internal registry where resources are described and connectors stored.

Policy management is ensured through the governance layer's policy stores.

Finally, the access control and identity management are the core functionalities of the SOI-SSG authorization service, identity broker and service gateway which enforces the authorisation decision identify users.

## VIII. CONCLUSION

In this paper, the authors have shortly introduced the concept of the Internet of Things & Services and linked it to the main software engineering domains it encompasses. Furthermore, the emphasis has been put on defining the requirements to secure resources in this large scale and dynamic paradigm. The key elements that have been identified are interoperability, automation, decentralisation and contextualisation.

Following this, the authors have proposed an architectural model, Security Cell, which takes these requirements into account and shortly described each of its components.

An example of a potential application of this architecture has been presented and defined in the SOI-SSG.

In future works, the authors will identify existing industrial products as well as research projects that attempt to address the issue of security in both this domain and related areas, and will analyse them against this model.